\documentclass[%
 aip,
 jmp,%
 amsmath,amssymb,
preprint,%
]{revtex4-1}

\usepackage{graphicx}% Include figure files
\usepackage{dcolumn}% Align table columns on decimal point
\usepackage{bm}% bold math
\usepackage[mathlines]{lineno}% Enable numbering of text and display math
\usepackage{subfigure}
\usepackage[para,online,flushleft]{threeparttable}
\usepackage{color}
\usepackage{dcolumn}
\usepackage[table,xcdraw]{xcolor}
\usepackage{longtable}
\usepackage{multirow}

\usepackage{booktabs}
\usepackage{amsmath}
\usepackage{multirow}
\usepackage{rotating}

\usepackage[toc,page,title,titletoc,header]{appendix}
\usepackage{amssymb,amsfonts,amsmath}

\usepackage{color, colortbl}
\definecolor{LightCyan}{rgb}{0.88,1,1}
\definecolor{Gray}{gray}{0.9}
\definecolor{Red}{rgb}{1.0, 0.22, 0.0}

\begin{document}

%\preprint{A Manuscript in Preparation}

\title{Frequency adaptive metadynamics for the calculation of rare-event kinetics}

%\thanks{A Manuscript in Preparation}

\author{Yong Wang}
\affiliation{
Structural Biology and NMR Laboratory, Linderstr{\o}m-Lang Centre for Protein Science, Department of Biology,
University of Copenhagen, Ole Maal{\o}es Vej 5 DK-2200 Copenhagen N, Denmark
}

\author{Omar Valsson}
\affiliation{Max Planck Institute for Polymer Research, Ackermannweg 10, D-55128 Mainz, Germany}

\author{Pratyush Tiwary}
\affiliation{Department of Chemistry and Biochemistry and Institute for Physical Science and Technology, University of Maryland, College Park 20742, USA.}

\author{Michele Parrinello}
\affiliation{Department of Chemistry and Applied Biosciences, ETH Zurich, c/o USI Campus, Via Giuseppe Buffi 13, CH-6900 Lugano, Switzerland}
\affiliation{Facolt\`{a} di Informatica, Instituto di Scienze Computationali, Universit\`{a} della Svizzera Italiana, CH-6900 Lugano, Switzerland}

\author{Kresten Lindorff-Larsen}%
 \email{lindorff@bio.ku.dk}
\affiliation{ 
Structural Biology and NMR Laboratory, Linderstr{\o}m-Lang Centre for Protein Science, Department of Biology,
University of Copenhagen, Ole Maal{\o}es Vej 5 DK-2200 Copenhagen N, Denmark
}%

\date{\today}% It is always \today, today,

\begin{abstract}
The ability to predict accurate thermodynamic and kinetic properties in biomolecular systems 
is of both scientific and practical utility. While both remain very difficult, 
predictions of kinetics are particularly difficult because rates, 
in contrast to free energies, depend on the route taken and are thus not amenable to all enhanced sampling methods. 
It has recently been demonstrated that it is possible to recover kinetics through so called 
`infrequent metadynamics' simulations, where the simulations are biased in a way that minimally corrupts the dynamics of moving between metastable states. 
This method, however, requires the bias to be added slowly, thus hampering applications to processes with only 
modest separations of timescales. 
Here we present a frequency-adaptive strategy which bridges normal and infrequent metadynamics. 
We show that this strategy can improve the precision and accuracy of rate calculations at fixed computational cost, 
and should be able to extend rate calculations for much slower kinetic processes.
\end{abstract}

%Use showkeys class option if keyword

\maketitle

%\begin{quotation}
%\end{quotation}

\section{Introduction}

Accessing long timescales in molecular dynamics (MD) simulations remains a longstanding challenge.
Limited by the short time steps of a few femtoseconds, and despite recent progress in methods and hardware \cite{Lindorff-Larsen2011,Buch2011,Tiwary2015a,Paul2017},
it remains difficult to access the timescales of milliseconds and beyond. This leaves many applications outside the realm of MD, including
many structural rearrangements in biomolecules and binding and unbinding events of ligands and drug molecules. Many of these reactions
are so called `rare events' where the time scale of the reaction is orders of magnitude longer than the time it takes to cross
the barrier between the states.
In such cases, most of the simulation time ends up being used in simulating the local fluctuations inside free energy basins.

To study phenomena that involve basin-to-basin transitions that occur on long timescales,
one typically has to rely on some combination of machine parallelism \cite{Dror2012,Lane2013},
advanced strategies for analysing simulations \cite{Prinz2011,Trendelkamp-Schroer2016,Chong2016} and enhanced sampling methods \cite{Bernardi2015,Valsson2016,DeVivo2016,Bruce2018}
that allow for efficient exploration of phase space.
Metadynamics is one such popular and now commonly used 
enhanced sampling method that involves the periodic application of a history-dependent biasing potential 
to a few selected degrees of freedom, typically also called collective variables (CVs) \cite{Laio2002}.
Through this bias, the system is discouraged from getting trapped in low energy basins,
and one can observe processes that would be far beyond the timescales accessible in normal MD,
while still maintaining complete atomic resolution.
Originally designed to explore and reconstruct the equilibrium free energy surface \cite{Bussi2015}, 
it has recently been shown that 
metadynamics can also be used to calculate kinetic properties
\cite{Tiwary2013,McCarty2015,Tiwary2015,Tiwary2015a,Mondal2016,Fleming2016,Tung2016,Sun2017,Fu2017,Wang2017,Tiwary2017,Casasnovas2017,Tiwary2017a}.
In particular, inspired by the pioneering work of Grubm{\"u}ller  \cite{Grubmueller1995} and Voter \cite{Voter1997}, 
it has recently been shown that the unbiased kinetics can be correctly recovered from metadynamics simulations
using a method called infrequent metadynamics (InMetaD)\cite{Tiwary2013}.

The basic idea in InMetaD is to add a bias to the free energy landscape sufficiently infrequently so that only the free energy basins, but not the free energy barriers,
experience the biasing potential, $V({s},t)$, where $t$ is the simulation time and $s$ is one or more CV's chosen to distinguish
the different free energy basins.
By adding bias infrequently enough compared to the time spent in barrier regions, the landscape can be filled up to speed up the transitions without perturbing the sequence of state-to-state transitions. The effect of the bias on time can then be reweighted through an acceleration factor:
\begin{equation}
\alpha(\tau)=\langle e^{\beta (V({s},t))} \rangle
\end{equation}
Here $\beta$ is the inverse temperature and the angular brackets denote an average over a metadynamics run
up until the simulation time $\tau$.

In the application of InMetaD to recover the correct rates, there are two key assumptions.
First, the state-to-state transitions are of the rare event type: 
namely, the system is trapped in a basin for a duration long enough
that memory of the previous history is lost, and when the system does translocate into another basin, 
it does so rather rapidly. 
Second, it is important that no substantial bias is added to the transition state (TS) region during the simulation.
This requirement can be achieved by adjusting the bias deposition frequency,
determined by the time between two instances of bias deposition.
If the deposition frequency is kept low enough, it becomes possible to keep the TS region unbiased.
This second requirement, however, also means that it takes longer to fill up the basin, revealing one
of the practical limitations of applying InMetaD.
It is this second requirement that we address and improve in this work.

In particular, we asked ourselves the question, can we design a bias deposition scheme so that
the frequency is high near the bottoms of the free energy basins,
but decreases gradually so as to lower the risk of biasing the TS regions?
Our scheme, which we term Frequency-Adaptive Metadynamics (FaMetaD),  is illustrated
using a ligand (benzene) unbinding from the T4 lysozyme (T4L) L99A mutant as an example (Fig. 1).
In particular, we designed a strategy that uses a high frequency at the beginning of the simulation (to fill up the basin quickly)
and then slows progressively down (to minimize the risk of perturbing the TS). 
In this way, we aim to improve the reliability, accuracy and robustness of the calculations
without  additional computational cost.

\begin{figure}[htbp]
 \begin{center}
\mbox{
{\includegraphics[height=6cm]{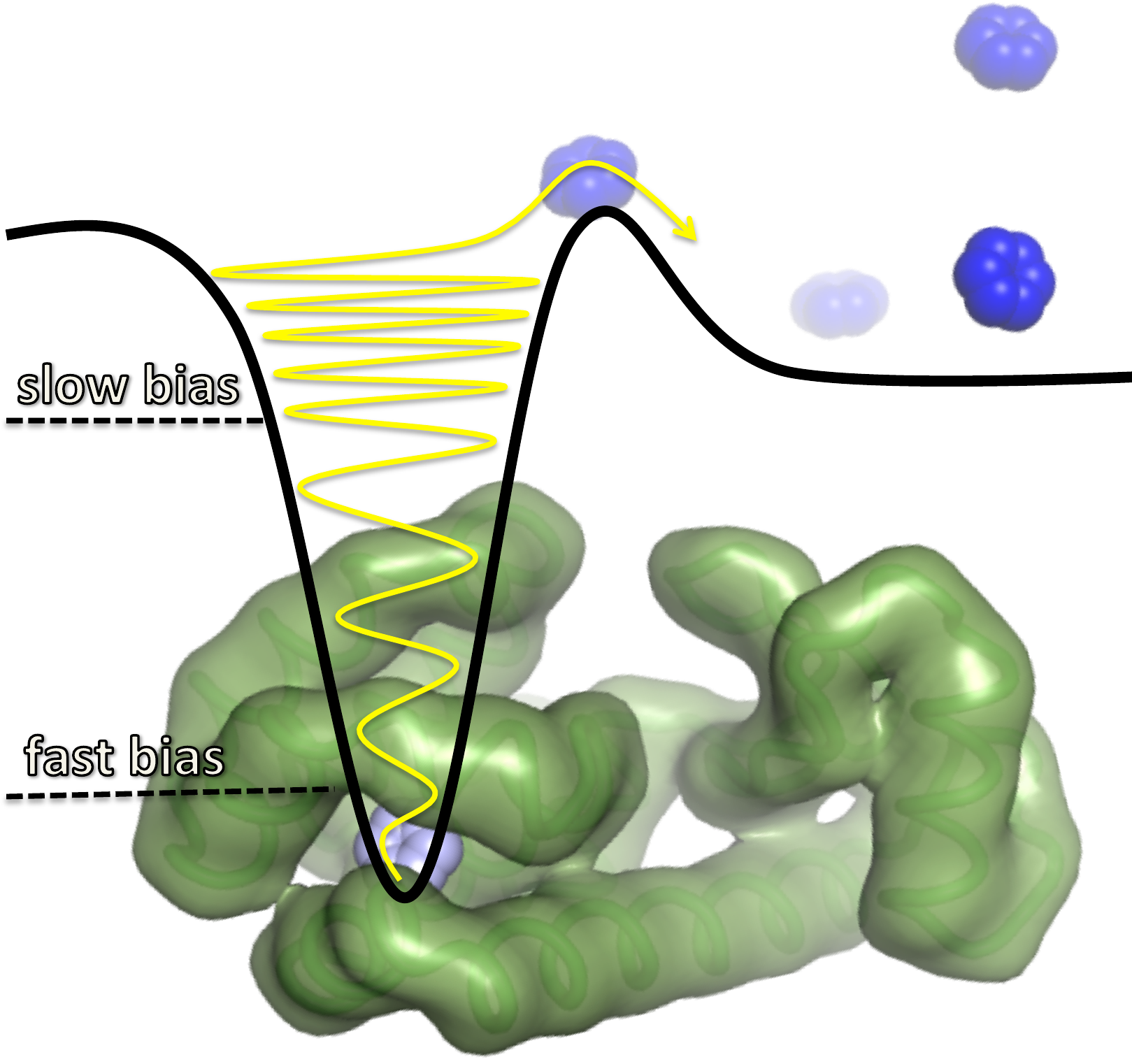}}
{\includegraphics[height=6cm]{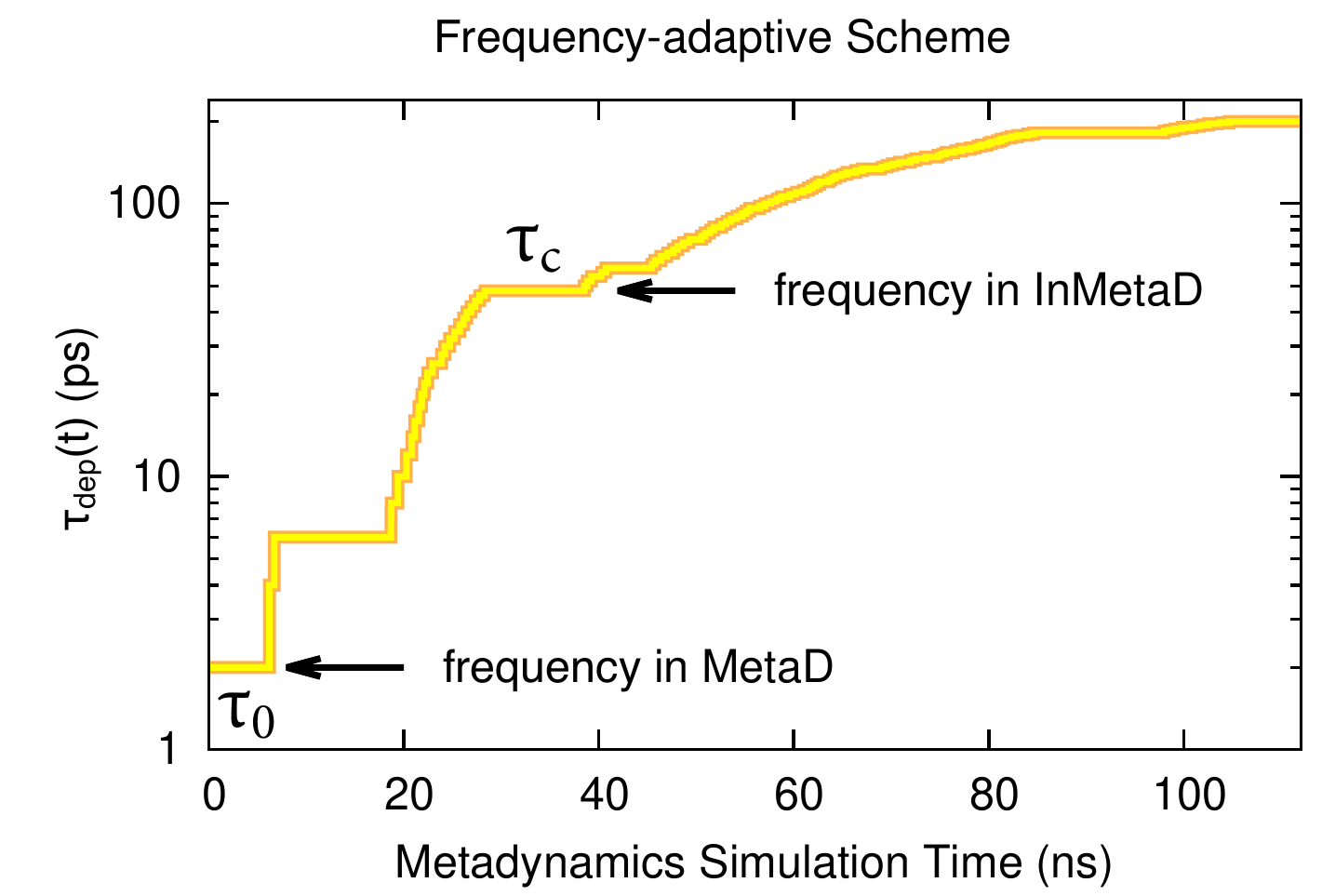}}
}
 \end{center}
 \caption{
{\bf A schematic picture to show the application of frequency-adaptive metadynamics (FaMetaD)
to calculate the off-rate of protein-ligand systems.}
The protein-ligand system (left panel)  is benzene (light and dark blue spheres) 
binding to the L99A mutant of T4 lysozyme (green cartoon) \cite{Wang2017}.
Frequency-adaptive metadynamics fills the free energy basin quickly at the beginning of the simulation,
but adds the bias slowly at the latter stage when the system moves close to the transition state regions.
The right panel shows a typical trajectory of how the time between deposition of the bias is adjusted on-the-fly
in a FaMetaD run.
The typical frequencies used in normal metadynamics ($\tau_0$) and InMetaD ($\tau_c$) are labeled by black arrows.
 }
\end{figure}

\section{Methods}
Our adaptive frequency scheme reads:
\begin{equation}
\tau_{dep}(t)=\min \{ \tau_0 \cdot \max\{\frac {\alpha(t)} {\theta},1\},\tau_c \}  
\end{equation}
where $\tau_0$ is the initial deposition time between adding a bias, similar to the relatively short time in normal metadynamics,
and $\tau_c$ is a cut-off value for the frequency equal to or larger than the deposition time
originally proposed in InMetaD.
These two parameters bridge the normal metadynamics and InMetaD,
by controlling the minimal and maximal bias deposition frequency.
$\alpha(t)$ is the instantaneous acceleration factor at simulation time $t$
in a metadynamics simulation.

Through the starting deposition time $\tau_0$, we can modulate the enhancement in our ability to fill free energy wells,
relative to an InMetaD run performed with a constant stride of $\tau_c$.
For practical considerations the choice of $\tau_c$ is dependent on the computational resources.
The key free parameter here is $\theta$, the threshold value for the acceleration factor 
to trigger the gradual change from normal (frequent) metadynamics to InMetaD.
The choice of $\theta$ requires some considerations.
On one hand, $\theta$ should be as large as possible to delay the switch from normal metadynamics
to InMetaD so as to get significant enhancement in basin filling.
On the other hand, too large $\theta$ could lead to problems 
that a transition might occur when the frequency $\tau_{dep}(t)$ is still less than 
the typical duration $\tau_c$ of a reactive trajectory crossing over from one basin to another.

We now estimate a value for $\theta$ in terms of the expected  transition time
$\tau_{exp}$ available from either experimental measurements or an  estimate,
the simulation time, $\tau_{sim}$ (determined also by computational resources), and a `safety coefficient' $C_s$.
We seek the following to hold true:
\begin{equation}
\theta \le \frac {\tau_{exp}} {\tau_{sim}C_s}.
\end{equation}
If we use a protein-drug system as an example, and we
(i) can estimate  the residence time to be roughly on the order of a second ($\tau_{exp}=1 s$), 
(ii) aim to observe the transition within a 100 ns metadynamics run ($\tau_{sim}=100 ns$), and
(iii)  use $C_s=10^2$ to counteract the risk that a transition time 
falls in the long tail of Poisson distribution, we thus obtain $\theta=\frac{1s}{100 ns * 10^2}=10^5$.
As we demonstrate by an example below, if one chooses too high a value of $\theta$ this leads to perturbed (and erroneous) kinetics;
an error that may be detected by testing whether the observed distribution of transition times follows a Poisson distribution  \cite{Salvalaglio2014}.

As per Eq. (2), the frequency is also changed as a function of $\alpha(t)$.
In practice, we observed significant fluctuation of $\tau_{dep}(t)$,
that  might cause problems if $\tau_{dep}(t)$ is small at the transition time point.
Therefore we finally modified Eq. (2) to become a monotonously increasing function: 
\begin{equation}
\tau_{dep}^{'}(t)=max(\tau_{dep}((N-1)\Delta t),\tau_{dep}(N \Delta t))
\end{equation}
where $\tau_{dep}(N\Delta t)$ is the instantaneous deposition time at step N
in a metadynamics simulation with a MD time step $\Delta t$.
The frequency-adaptive scheme was implemented in a development version of the PLUMED2.2 code\cite{Tribello2014}.

\begin{table*}
%\begin{table}[ht]
\caption{\bf Conformational Transition Times of Ace-Ala3-Nme}
\begin{center}
 \begin{threeparttable}
  \begin{tabular}{c ccc c c}
%\toprule
\hline
\rowcolor{Gray}
                        & Parameters: $\tau_0$ (ps), $\tau_c$ (ps), h (kJ/mol)           & $\tau_{slow}$ ($\mu s$) & P-value       & Cost ($\mu$s) &  Set \\
\hline
Unbiased MD             & T=300K                                & 11$\pm$2\tnote{a}      &             & 300 & \\
InMetaD                 & $\tau_0=200$,$h=0.4$                  & 16$\pm$5               & 0.4$\pm$0.3 & 2.8 &   \\
                        & $\tau_0=100$,$h=0.4$                  & 12$\pm$3               & 0.3$\pm$0.2 & 1.5 & A \\
                        & $\tau_0=40$,$h=0.4$                   & 19$\pm$6               & 0.3$\pm$0.2 & 0.8 & B \\
                        & $\tau_0=20$,$h=0.4$                   & 16$\pm$5               & 0.1$\pm$0.2 & 0.5 & C \\
                        & $\tau_0=10$,$h=0.4$                   & 32$\pm$15              & 0.1$\pm$0.1 & 0.3 & D \\
                        & $\tau_0=5$,$h=0.4$                    & 92$\pm$38              & 0.02$\pm$0.05 & 0.2 & E \\
                        & $\tau_0=2$,$h=0.4$                    & 98$\pm$61              & 0.01$\pm$0.01 & 0.1 & F \\
FaMetaD              & $\tau_0=2$,$\tau_c=200$,$\theta=1$,$h=0.4$       & 15$\pm$3     & 0.4$\pm$0.3 & 1.5 & A  \\
                        & $\tau_0=2$,$\tau_c=200$,$\theta=10$,$h=0.4$   & 14$\pm$3     & 0.5$\pm$0.3 & 0.6 & B  \\
                        & $\tau_0=2$,$\tau_c=200$,$\theta=40$,$h=0.4$   & 19$\pm$6     & 0.2$\pm$0.2 & 0.4 & C  \\
                        & $\tau_0=2$,$\tau_c=200$,$\theta=100$,$h=0.4$  & 24$\pm$7     & 0.1$\pm$0.1 & 0.3 & D  \\
                        & $\tau_0=2$,$\tau_c=200$,$\theta=500$,$h=0.4$  & 29$\pm$10    & 0.1$\pm$0.2 & 0.2 & E  \\
                        & $\tau_0=2$,$\tau_c=200$,$\theta=10000$,$h=0.4$   & 24$\pm$11    & 0.02$\pm$0.03 & 0.1 & F \\
\hline
  \end{tabular}
  \begin{tablenotes}[para,flushleft]
  \footnotesize
   \item [a] From ref. \cite{Wang2017}.
%  \item [a] We collected 28 successful transitions from 50 trials, the maximum likelihood estimate of the transition time is calculated by
%$\tau_{MLE}=\frac {\sum_{i=1}^n t_i + \sum_{j=1}^{N-n} t_j} {n} \pm \frac {\sum_{i=1}^n t_i + \sum_{j=1}^{N-n} t_j} {b^{3/2}} $ \cite{Zagrovic2003}.
  \end{tablenotes}
 \end{threeparttable}
\end{center}
\end{table*}

\section{Results and Discussion}
\subsection{Test on a four-state model system}

To benchmark our results, we first consider a five-residue peptide (Nme-Ala3-Ace) as a model system
with a non-trivial free energy landscape involving multiple conformational states \cite{Wang2017}.
We consider the slowest state-to-state transition time, $\tau_{slow}$, 
which has been estimated to be $\sim 11 \mu s$ from unbiased MD \cite{Wang2017}, as the benchmark target.
We used the same computational setup and CVs as previously described \cite{Wang2017}.
%In this example a suitable choice for the collective variable is the fcc order parameter introduced in ref 52. 

As a baseline, we used InMetaD
with a fixed height of Gaussian bias potential ($h=0.4$ kJ/mol) 
but with different times of bias deposition ranging from 2 ps to 200 ps (40 runs for each parameter set)
(Table 1 and  Fig. S1).
The reliability of the calculated transition times were verified a posteriori using a Kolmogorov-Smirnov test \cite{Salvalaglio2014}
to examine whether their cumulative distribution function is Poissonian.
If the $p$-value is low (e.g. less than 0.05) then it is likely that the distribution has been perturbed
because of too aggressive application of the biasing potential.
As expected, we find that simulations that used a low frequency ($\tau_0\ge20$ ps) 
gave rise to a consistent estimation of $\tau_{slow}$,
with values close to those observed in unbiased MD,
while the simulations with high frequency ($\tau_0\le10$ ps) resulted in substantially longer 
and less reliable times (Table 1 and  Fig. S1).
This correlation between the $p$-value and the deposition time in the InMetaD simulations can be explained by the fact that 
the higher bias frequency results in higher risk of biasing the TS regions and more non-Poissonian distribution.
We also note that the transition times appear to be over-estimated in these cases.

To compare the FaMetaD simulations with the InMetaD baseline,
we designed six sets (A to F) of FaMetaD simulations to have comparable computational costs (Table 1 and Fig. 2).
Overall, the results of FaMetaD show the same trends as InMetaD and support the same conclusions.
In other words, both InMetaD and FaMetaD reveal the trade-off between
reliability, accuracy and computational cost. There are, however, some important differences 
that highlight the improvement obtained through FaMetaD. First, in each case there is a small but notable improvement
in the reliability of the calculations, as judged by the greater $p$-values that suggest that FaMetaD perturbs the TS less than
InMetaD. Most important, however, is the accuracy of the calculations. While both InMetaD and FaMetaD achieve accurate results
when using the most conservative parameters (sets A--C), there are dramatic differences when more aggressive parameters are used (D--F).
Importantly, we observe a much more `gracious' decline in accuracy with more aggressive parameters in FaMetaD comparing to InMetaD.
Thus together these results suggest that
the frequency-adaptive scheme can further improve not only the reliability but also the accuracy of the calculation,
without the need of increasing computational burden, and also that the reliability is more robust to the choice of the simulation parameters.

\begin{figure}[htbp]
 \begin{center}
\mbox{
{\includegraphics[width=12cm]{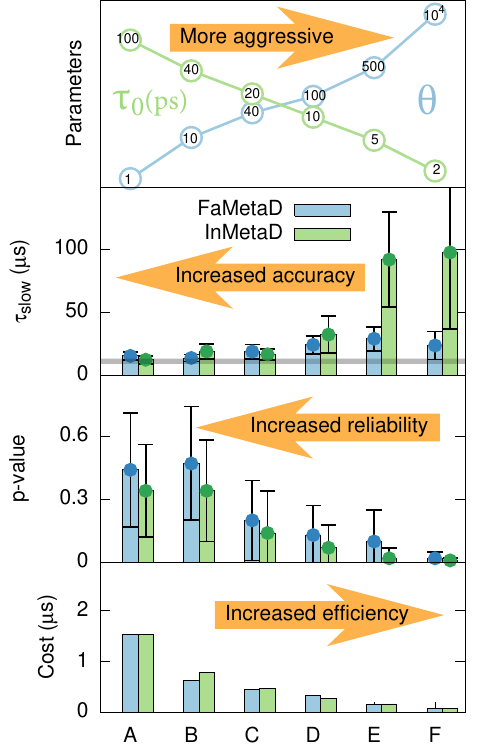}}
}
 \end{center}
 \caption{
{\bf Comparing the accuracy, reliability and efficiency of frequency-adaptive metadynamics 
and infrequent metadynamics.}
The upper panel shows the key parameters ($\tau_0$ in InMetaD, $\theta$ in FaMetaD) of the six sets (A to F) pairs simulations.
The middle panels show a comparison of the accuracy and reliability of the results.
The grey line shows the $\tau_{slow}$ obtained from unbiased MD simulations.
The bottom panel shows the computational cost of each set of simulations.
 }
\end{figure}

\subsection{Application on protein-ligand binding}

\begin{table*}[ht]
\caption{\bf Binding and unbinding times of T4L L99A with benzene and indole}
\begin{center}
 \begin{threeparttable}
  \begin{tabular}{c cccc}
\hline
Methods   & Parameters: $\tau_0$ (ps), $\tau_c$ (ps), $h$ (kJ/mol)              & Time (ms) & p-value       & Cost     \\ 
\hline
\rowcolor{Gray}
            \multicolumn{5}{c}{Set 1: Benzene Binding ($\tau_{on}^{BNZ}$)}           \\
InMetaD\tnote{b}   &  $\tau_0=40$,$h=0.4$                   & 9$\pm$5              & 0.1$\pm$0.1 & 4.4$\mu$s \\%(221ns/run)  \\
FaMetaD & $\tau_0=1$,$\tau_c=100$,$h=0.2$,$\theta=10^3$   & 14$\pm$7             & 0.2$\pm$0.1 & 3.0$\mu$s \\%(169ns/run)  \\
\rowcolor{Gray}
            \multicolumn{5}{c}{Set 2: Benzene Unbinding ($\tau_{off}^{BNZ}$)}        \\
InMetaD\tnote{b}   &  $\tau_0=100$,$h=0.2$                  & 168$\pm$59           & 0.4$\pm$0.3 & 6.7$\mu$s \\%(334ns/run)  \\
FaMetaD & $\tau_0=1$,$\tau_c=100$,$h=0.2$,$\theta=10^3$   & 176$\pm$68           & 0.3$\pm$0.2 & 5.5$\mu$s \\%(274ns/run)  \\
\rowcolor{Gray}
            \multicolumn{5}{c}{Set 3: Indole Unbinding ($\tau_{off}^{IND}$)}         \\     
InMetaD   & $\tau_0=100$,$h=0.2$                   & 102$\pm$87           & 0.2$\pm$0.2 & 4.5$\mu$s \\%(298ns/run)  \\
FaMetaD & $\tau_0=1$,$\tau_c=100$,$h=0.2$,$\theta=10^4$   & 168$\pm$95           & 0.2$\pm$0.1 & 2.1$\mu$s \\%(138ns/run)  \\
\hline
        & & & & 26$\mu$s \\
\hline
  \end{tabular}
  \begin{tablenotes}[para,flushleft]
  \footnotesize
  \item [a] In all simulations, the ligand concentration is $\sim 5$ mM. 
  \item [b] The results are from our recent work \cite{Wang2017}.
  \end{tablenotes}
 \end{threeparttable}
\end{center}
\end{table*}

We consider now the more complex case of a ligand binding to or escaping from a buried internal cavity
in the L99A mutant of T4 lysozyme, processes which occur on timescales of milliseconds or more
\cite{Feher1996,Bouvignies2011,Wang2016}.
To benchmark the results, we performed three sets of metadynamics simulations, up to 26$\mu$s in total 
(including 12 $\mu$s InMetaD simulations of T4L L99A with benzene from our recent work \cite{Wang2017}).
In each set, we compared the results of InMetaD with that of FaMetaD.
In set 1 and 2 we calculated the time constants for binding ($\tau_{on}^{BNZ}$) and unbinding ($\tau_{off}^{BNZ}$) of benzene,
while we in set 3 calculated the time for indole to escape the pocked ($\tau_{off}^{IND}$).
We performed 20 independent runs for each set of simulations (using the CHARMM22* force field \cite{Piana2011} for protein
and CGenFF force field \cite{Vanommeslaeghe2010} for the ligands)
to collect the transition times, from which we obtained $\tau_{on}$ and $\tau_{off}$ (Table 2).

Again, we find consistency between the results of InMetaD and FaMetaD, with e.g. $\tau_{on}^{BNZ}$ and $\tau_{off}^{BNZ}$ to be $\sim 10$ ms and $\sim 170$ ms, respectively.
As previously described \cite{Wang2017}, we can use these values
to estimate the binding free energy to be $\Delta G_{binding}\approx -21$ kJ/mol, a value that agrees well with 
calorimetric \cite{Morton1995} and NMR \cite{Feher1996} measurements.

In set 1, the InMetaD simulation was performed with $\tau_0=40$ ps and $h=0.4$ kJ/mol.
The frequency-adaptive scheme allows us to perform FaMetaD with weaker bias ($h=0.2$ kJ/mol)
and ending with longer and more conservative deposition time ($\tau_c=100$ ps).
This parameter set used less simulation time but resulted in a more reliable estimation.
%(with p-value of $0.2\pm0.1$ in FaMetaD vs $0.1\pm0.1$ in InMetaD).
In set 2, the FaMetaD simulation was performed using the same $\tau_c=100$ ps and $h=0.2$ kJ/mol 
as that used in the InMetaD simulation, but with $\theta=10^3$.
Given that $\tau_{exp}\sim 10-100$ ms, $\tau_{sim}\sim 100$ ns and $C_s=10^2$,
according to Eq. (3), this allows us to judge that $\theta=10^3$ is a fairly conservative parameter.
The estimated values of $\tau_{off}^{BNZ}$ in this set are indistinguishable 
between InMetaD and FaMetaD within the error bars but at slightly less computational cost.
In set 3, we used similar parameters as in set 2 except a bit more aggressive $\theta=10^4$.
Again, this parameter set resulted in $\tau_{off}^{IND}$ of $168\pm95$ ms from FaMetaD
that is reasonably close to the estimation from InMetaD.
Remarkably, FaMetaD allowed us to reduce more than half the computational cost,  without loss of accuracy.
Overall, the application in the case of L99A binding with two ligands suggests that the frequency-adaptive scheme
can improve the reliability-accuracy-efficency balance of metadynamics on kinetics calculation.

\section{Conclusions}
Many biological processes occur far from equilibrium, and  kinetic properties can play an important role
in biology and biochemistry. 
For example, there has been continued interest in determining ligand residence times  in the context of drug
optimization \cite{Copeland2006}, 
and millisecond conformational dynamics in an enzyme has been shown to underlie an intriguing
phenomenon of kinetic cooperativity of relevance to disease \cite{Larion2015}.

The principle of microscopic
reversibility, however, sets limits to how the kinetic properties can be varied independently of thermodynamics. Thus, for example,
increasing the residence time of a ligand will also increase its thermodynamic affinity, unless there is also
a simultaneous drop in the rate of binding. Thus, in practice it may be in many cases be difficult to disentangle
the effects of kinetics and thermodynamics.

Taken together, the above considerations suggest the need for improved methods for understanding and ultimately predicting
the rate constants of biological processes. Further, the ability to calculate the kinetics of conformational exchange or ligand binding and unbinding
provides an alternative approach to determine equilibrium properties \cite{Wang2017}.
For these and other reasons, several methods have been developed to enable the estimation of kinetics from simulations \cite{Bruce2018}.

We have here proposed a modification to the already very powerful InMetaD algorithm, which leads to a further improvement 
in the reliability and accuracy in recovering unbiased transition rates from biased metadynamics simulations.
The basic idea in the resulting FaMetaD approach is that, by filling up the basin more rapidly in the beginning and only
using an infrequent bias near the barrier, we may spend the computational time where it is needed.
We anticipate that our scheme will prove particularly useful in two different aspects. 
First, by enabling a decreased bias in the transition state region,
we obtain more accurate kinetics at fixed computational cost, leading also to the observed robustness of our approach. Second, in the case of large barriers,
it would be prohibitively slow to use a very infrequent bias through the entire duration of the simulation. Thus, the FaMetaD provides a practical approach
to study  rare events that involve escaping deep free energy minima, and which occur on timescales well beyond those accessible to current simulation methods.
Here we have opted to examine processes that can be studied using both InMetaD and FaMetaD, but in the future we aim to apply this approach to barrier crossing
events that occur on even longer timescales. With more examples in hand, we also expect that in the future
it may be possible to design improved biasing schemes to increase the computational
efficiency further, and at the same time retain the robustness of the FaMetaD approach.

\section{Acknowledgement}
The authors thank Tristan Bereau and Claudio Perego for a critical reading of the manuscript.
K.L.-L. acknowledges funding by a Hallas-M{\o}ller Stipend from the Novo Nordisk Foundation
and the BRAINSTRUC initiative from the Lundbeck Foundation.
M.P. acknowledges funding from the National Centre for Computational Design and Discovery of Novel Materials MARVEL and European Union grant ERC-2014-AdG-670227/VARMET.

%\bibliography{../PaceRef}
%

\end{document}